\documentclass[USenglish]{article}
\usepackage[utf8]{inputenc}
\usepackage{graphicx}%
\usepackage[font=small,labelfont=bf]{caption} 
\usepackage{amsfonts, amsmath, amsthm, amssymb} 
\usepackage{wrapfig} 
\usepackage[version=3]{mhchem}
\usepackage{verbatim, soul}
\usepackage{mathtools,authblk}
\usepackage{caption, subcaption}
\usepackage{lmodern}
\RequirePackage{fix-cm}
\usepackage{accents}

\newcommand{\jq}{{j}_q}
\newcommand{\js}{{j}_s}
\newcommand{\sigmas}{\sigma_s}

\DeclareMathOperator*{\argmax}{arg\,max}
\DeclareMathOperator*{\argmin}{arg\,min}

\newcommand{\ha}{h_\alpha}
\newcommand{\he}{h_\text{e}}
\newcommand{\sa}{s_\alpha}
\newcommand{\se}{s_\text{e}}

\newcommand{\phie}{\varphi_{\rm e}}
\newcommand{\phe}{\phi_\text{e}}
\newcommand{\phii}{\phi_{\rm i}}

\newcommand{\ja}{{j}_\alpha}

\newcommand{\Pitot}{\Pi_{\rm tot}}
\newcommand{\Kh}{K_{\rm h}}
\newcommand{\Ko}{K_{\rm o}}

\newcommand{\Aft}{\widetilde{\mathcal{A}}}
\newcommand{\Aftot}{\Aft_{\rm tot}}

\newcommand{\dksi}{\dot{\xi}}

\newcommand{\je}{{j}_\text{e}}
\newcommand{\jqp}{{j}_{q}'}
\newcommand{\jqb}{{j}_{\rm qb}'}

\newcommand{\dWel}{\dot{W}_{\rm el}}

\newcommand{\dd}[2]{\frac{{\rm d}{#1}}{{\rm d}{#2}}}

\begin{document}

\title{Pitfalls of exergy analysis}

\author[1,2]{Petr V\'agner}
\author[2,3]{Michal Pavelka}
\author[1]{Franti\v sek Mar\v s\'ik} 

\affil[1]{\protect\raggedright 
  New Technologies - Research Centre, University of West Bohemia, Univerzitní 8, 306 14 Pilsen, Czech Republic
  , e-mail: marsik@it.cas.cz}
\affil[2]{\protect\raggedright 
Mathematical Institute, Faculty of Mathematics and Physics, Charles University in Prague, Sokolovská 83, 186 75 Prague, Czech Republic, e-mail: affro@atrey.karlin.mff.cuni.cz, michal.pavelka@email.cz}
\affil[3]{\protect\raggedright Department of Chemical Engineering, University of Chemistry and Technology Prague, Technick\'a 5, 16628 Prague 6, Czech Republic}
	
\maketitle
\abstract{The well-known Gouy-Stodola theorem states that a device produces maximum useful power 
when working reversibly, that is with no entropy production inside the device. This statement 
then leads to a method of thermodynamic optimization based on entropy production minimization. 
Exergy destruction (difference between exergy of fuel and exhausts) is also given by entropy production inside 
the device. Therefore, assessing efficiency of a device by exergy analysis is also based on the 
Gouy-Stodola theorem. However, assumptions that had led to the Gouy-Stodola theorem are not satisfied 
in several optimization scenarios, e.g. non-isothermal steady-state fuel cells, 
where both entropy production minimization and 
exergy analysis should be used with caution. We demonstrate, using non-equilibrium thermodynamics, 
a few cases where entropy production minimization and exergy analysis should not be applied. \textit{Accepted to Journal of Non-equilibrium Thermodynamics 2016.}
}



\section{Introduction}
In 1889 Gouy published paper \cite{Gouy}, where he showed how to calculate useful power of a device by thermodynamic means. The calculation was based on two assumptions: (i) The environment surrounding the device is isothermal (ambient temperature) and (ii) the mechanical power should be maximized. Similar result was obtained by Stodola in \cite{Stodola}. Over more than a century of development was comprehensively reviewed in \cite{Bejan1996}, where also the method of entropy production minimization (EPM) was elucidated. An advantage of thermodynamic optimization based on EPM is that one can plot a continuous map of losses (given by entropy production within EPM) revealing how much efficiency is lost at each place of the device, see e.g. \cite{Sciacovelli2010}.

To obtain the continuous map of losses, which is often proportional to entropy production density, one has to solve the continuum non-equilibrium thermodynamic equations governing the system under consideration. Usually the system is in a steady-state (not evolving in time), and classical irreversible thermodynamics (CIT), developed in \cite{Meixner,dGM}, in the form presented by Bedeaux and Kjelstrup, \cite{Kjelstrup,Kjelstrup-Engineers}, provides a systematic approach for developing thermodynamic descriptions of the systems.

An alternative to continuum non-equilibrium thermodynamics is the endoreversible thermodynamics  \cite{Hoffmann97,Sieniutycz1998,Hoffmann03} or finite-time thermodynamics \cite{Curzon-Ahlborn,Novikov}, where some parts of the system are studied in detail (as in the continuum approach) while some are described only on the macroscopic level of equilibrium thermodynamics as in the theory behind the Gouy-Stodola theorem. Such approach is advantageous in engineering applications because it reduces the amount of detail required in the full continuum calculations.
 
However, before trying to plot a map of losses, it is necessary to define what the losses mean in terms of the state variables chosen for description of the system, e.g. fields of concentrations, temperature and electric potential. Such task inevitably leads to the choice of an objective function that is to be maximized and constraints that are to be kept constant during the maximization. Regarding the Gouy-Stodola theorem, it might seem natural to identify the losses with entropy production, but it has been shown for example in \cite{Salamon-Hoffmann-Andersen, Pavelka2013, GenExAl} that it is not so quite often. Firstly, one can choose a different objective function than electric power, in which case entropy production clearly does not need to describe the losses. Secondly, which is more important, even if one chooses useful work as the function that should be maximized, entropy production is often inadequate measure of losses for example when boundary of the system is not isothermal.

Indeed, it was shown in \cite{Pavelka2013} and \cite{GenExAl} that when considering a non-isothermal fuel cell in steady state, entropy production is inadequate to address the map of losses of electric power, since electric power is given by the flux of Gibbs energy into the fuel cell diminished by a functional different from entropy production. In other words, consider a steady state non-isothermal fuel cell with some flux Gibbs energy into the fuel cell. What is the maximum electric power one can produce? The maximum work coincides with minimum of a functional that is definitely different from entropy production. 

So, maximum work at steady state does not necessarily correspond to minimum entropy production, which is zero entropy production by the second law of thermodynamics. Consider now the difference in exergy of the fuel and exhausts coming into and out of the fuel cell. This loss of exergy is often used as a measure of losses in the device, see for example review article \cite{Bejan1996}, \cite{Arshad2010} for diffuser optimization, \cite{AlSulaiman2013} for humidification-dehumidification system optimization, \cite{Gutierrez2012} for methane decomposition optimization, \cite{Ishida1987} for combustion optimization, or \cite{Sciacovelli2016} for thermal storage optimization. But exergy destruction is proportional to entropy production as shown in \cite{Bejan1996}, and since entropy production is often inadequate to measure losses in the device, analysis of exergy must be also often inapplicable, if power maximization is the optimization goal.%

One could argue that exergy represents the maximum work one can obtain from the device when the device relaxes to equilibrium with its surroundings. This is of course true as shown for example in \cite{Landau5}, \S 20. But is it always the goal of exergy analysis to find how much work the device could deliver when relaxing to equilibrium, for example when shutting down a power plant? 

In summary, before one decides to measure efficiency of a device or a component of the device by evaluating exergy destruction, one should either declare that the goal is to find the maximum power the device can deliver when relaxing to thermodynamic equilibrium, which means also shutting down the device, or one should verify that zero entropy production corresponds to the most efficient steady state of the device. The former is relatively simple, but restrictive, since the device often works in a steady state. The latter needs a clear definition of efficiency (or an objective function, e.g. electrical power) and constraints, and usually relies on using continuum non-equilibrium thermodynamics.

Let us illustrate the latter approach on a fuel cell in a steady state, as in \cite{GenExAl}. The objective function is the electrical power. The constraints contain flux of Gibbs energy into the fuel cell. It follows from non-equilibrium thermodynamics that if boundary of the fuel cell is isothermal, maximization of power is equivalent to minimization of entropy production inside the fuel cell. If the boundary is not isothermal, the equivalence is lost, and one should minimize a different functional than entropy production, which means that exergy destruction (which is proportional to entropy production) is inadequate to measure efficiency of the fuel cell. 

It is the purpose of this paper to shed more light on such pitfalls one can meet when performing thermodynamic optimization. A one-dimensional steady state fuel cell is considered and described within non-quilibrium thermodynamics in the form of \cite{Kjelstrup}. The model is chosen so that it can be solved analytically. The objective function is the electrical power and several examples of constraints are then considered. It is demonstrated that entropy production minimization does not coincide with power maximization. Finally, several examples are identified where exergy analysis and entropy production minimization are appropriate tools of power maximization. We hope that readers will be discouraged from blind using of exergy analysis.
\section{Global balance laws}
Consider a one-dimensional thermodynamic system, for example a fuel cell, in nonequilibrium steady-state. 
Although the total energy of the system is constant in time (steady state), there is non-zero flux of energy through boundary of the system, as well as non-zero fluxes of particular species (fuel, exhausts, and electrons). These transport processes are accompanied by transport of entropy. There is no source of energy in the system, which means that energy fluxes through the boundary have to sum up to zero. On the other hand, entropy is being produced inside the fuel cell, which means that fluxes through the boundary sum up to the total entropy production inside the fuel cell. 

Boundary of the system is characterized by two points, $0$ and $L$. Difference of any quantity between the two points will be denoted by $\Delta(\bullet) = \bullet^L - \bullet^0$. It is assumed that electrochemical reactions take place at the boundary, i.e. within the two points, and so the two points themselves are also equipped with balance equations.

\subsection{Total energy balance}
Balance of total energy of the system reads{\footnote{The partial enthalpies $\ha$ differ 
from partial enthalpies of pure components, for details see Appendix B in \cite{Pavelka2014}.}}
\begin{equation}\label{eq:sysEnBal1}
\frac{\partial}{\partial t} {E_\text{total}} = 0 = \Delta \jq + \sum_\text{neutral}%
\Delta(\ja\ha) + \Delta\big(\je(\he + \phie)\big).%
\end{equation}
Subscript $\alpha$ denotes association to species $\alpha$, in particular subscript $\text{e}$ is reserved to denote electrons.
Symbols $j_\alpha$ and $h_\alpha$ stand for molar flux of species $\alpha$ and partial molar enthalpy of the species, respectively. 
Symbol $\phie$ stands for energy of electrons due to electrostatic field $\varphi$, i.e. $\phie = - \varphi$, and symbol $\jq$ denotes heat flux.

From the practical point of view we cannot distinguish between chemical potential of charged species, $\mu_\alpha$, and the electrostatic potential energy of the species, $z_\alpha F \varphi${\footnote{\label{note2} $z_\alpha$ stands for elementary charge per particle of species $\alpha$ and $F$ stands for Faraday constant.}}. Therefore, we prefer working with electrochemical potential, $\tilde{\mu}_\alpha = \mu_\alpha + z_\alpha F \varphi$, and define electric potential of charged species $\alpha$ as 
\begin{equation}
z_\alpha F \phi_\alpha : = \mu_\alpha + z_\alpha F \varphi,
\end{equation}
which was proposed for example in \cite{Kjelstrup}, where the electrostatic potential, $\varphi$, is referred to as Maxwell potential. In particular, electric potential is defined as the electric potential of electrons,
\begin{equation}\label{eq:electron_potential}
- F \phi_\text{e} = \mu_e - F \varphi.
\end{equation}
This is indeed the quantity measured by a voltmeter, since voltmeter in fact measures the tiny current passing through it, which deflects the arrow of the voltmeter by electromagnetic induction, and the current is proportional to difference in electrochemical potential of the electrons across the voltmeter. See \cite{My-baterky} for more discussion.

Electrical work produced by the system can be expressed as the energy flux due to 
electrons passing boundary of the system,{\footnote{\label{note1} For charged species we define $i_\alpha := z_\alpha F \ja$, in
particular, we have $i_{\rm e} := - F \je$ for electrons.}} 
\begin{equation}
 \dWel = \Delta (i_e \phi_\text{e}),
\end{equation}
where $i_e$ is electric current density due flux of electrons. Similarly, $i_\alpha$ is electric current density due to a charged species $\alpha$. Using the relation between enthalpy, chemical potential and entropy 
\begin{equation}
h_\alpha = \mu_\alpha + T s_\alpha,
\end{equation}
we obtain from \eqref{eq:sysEnBal1} that
\begin{equation}\label{eq:sysEnBal2}%
\dWel =%
			-\Delta(\underbrace{\jq + T\je\se}_{\jqp})%
			- \sum_{\mathclap{\text{neutral}}}\Delta(\ja\ha).%
\end{equation}%
which is an another form of total energy balance \eqref{eq:sysEnBal1}. Note the usage of measurable heat flux $\jqp$, introduced in \cite{Kjelstrup}. 
The measurable heat flux helps to keep the energy balance free of the electron entropy flux, which we cannot measure
experimentally anyway. Equation \eqref{eq:sysEnBal2} contains the usual meaning of balance of energy, that electrical work is equal to heat and enthalpy flux into the system. 

\subsection{Entropy balance}
As in the case of total energy, entropy cannot accumulate inside the system due to steadiness of the state.
Unlike energy, entropy is produced inside the considered system due to nonequilibrium nature of the state. 

Flux of entropy and entropy production can be expressed as
\begin{subequations}
\begin{eqnarray}
 \js &=& \frac{\jqp}{T} + \sum_\text{neutral}\ja\sa,\\
 \sigma_s &=& \jqp \cdot \frac{\partial}{\partial x} \frac{1}{T} 
  - \frac{1}{T} \sum_{\mathclap{\text{charged}}}i_\alpha \cdot \frac{\partial \phi_\alpha}{\partial x} 
 - \frac{1}{T} \sum_{\mathclap{\text{neutral}}}j_\alpha \cdot \left(\frac{\partial \mu_\alpha}{\partial x}\right)_T\nonumber\\
 &&+\frac{1}{T} \sum_r \widetilde{A}_r \dot{\xi}_r.
\end{eqnarray}
\end{subequations}
Gradient of chemical potential at constant temperature,
\begin{equation}
\big(\frac{\partial \mu_\alpha}{\partial x}\big)_T = \frac{\partial \mu_\alpha}{\partial x}-\underbrace{\frac{\partial\mu_\alpha}{\partial T}}_{=-s_\alpha}\frac{\partial}{\partial x} T,
\end{equation}
serves as a driving force for uncharged species. We considered that (electro)chemical reactions are taking place among the species. Electrochemical affinity of reaction $r$ is defined as
\begin{equation}\label{eq:defAft}
 \widetilde{A}_r = -\sum_\alpha \nu^r_\alpha \tilde{\mu}_\alpha
\end{equation}
where $\nu^r_\alpha$ is the stoichiometric coefficient of species $\alpha$ in reaction $r$. Rate of the reaction (in $mol/m^3 s$) is denoted by $\dot{\xi}_r$. See \cite{GenExAl} or \cite{Kjelstrup} for derivation of these formulas.

Finally, flux of entropy out of the system is equal to the total entropy production inside the system, which means that 
\begin{equation}\label{eq:sysEpBal1}
\Delta\Big(\frac{\jqp}{T} + \sum_\text{neutral}\ja\sa\Big) = \int  \sigma_s \mathrm{d}x\stackrel{\mbox{def}}{=}\Pi.
\end{equation}
Total entropy production inside the system is denoted by $\Pi$. Note that the electron entropy flux
is integrated into the measurable heat flux.
Second law of thermodynamics asserts general positiveness of entropy production, which implies that
\begin{equation}
\Pi \geq 0.
\end{equation}
\subsection{Efficiency of a device producing electricity}
Consider a device producing electrical work by converting heat or chemical energy into electric energy, e.g. a
hydrogen fuel cell. Plugging $\jqp^0$ from the entropy balance \eqref{eq:sysEpBal1} into energy balance \eqref{eq:sysEnBal2} yields
\begin{equation}\label{eq:sysEnBalEp1}
{\dWel} =%
			-\jqp^L\left(1 - \frac{T^0}{T^L} \right)%
			+ \sum_\text{neutral}\Delta\left(\ja(T^0\sa - \ha)\right)%
			-T^0\Pi.
\end{equation}
This last equation connects electric power and entropy production.
There is only one term on the right hand side which is always non-positive, $-T^0\Pi$. 
Let us assume that the objective function we wish to maximize is the electrical work.
Then it seems natural to design the device so that the entropy production is minimal while keeping the
resources, the first two terms on the right hand side of Eq. \eqref{eq:sysEnBalEp1}, constant, which leads to the method of Entropy Production Minimization (EPM), reviewed in \cite{Bejan1996}. The resources are equal to exergy flux into the device and the non-positive term is negative of the exergy destruction. It is clearly true that when keeping exergy flux constant, the useful work is maximal when exergy destruction (or entropy production) is minimal.

Consider non-isothermal boundary of the system. Exergy flux into the system then contains heat fluxes through all parts of the boundary except for the part with temperature $T_0$, temperature reservoir $T_0$. This temperature reservoir is often referred to as the surroundings. What if we do not wish to keep all those heat fluxes constant when performing the maximization? That is often the case for example in fuel cells, where efficiency is expressed as the ratio of electrical work and flux of Gibbs energy into the system. No heat fluxes appear in the definition of efficiency, and thus one could seek for maximum work when fixing only flux of Gibbs energy into the fuel cell. 

Such choice of optimization constraints has the important implication that, since exergy flux is no longer constant, entropy production is no longer the functional that should be minimized. See \cite{GenExAl} for quantitative results. This idea is further explored in the rest of this paper.
\subsection{Physical model, constraints and optimization}
\label{subsec:OptGen}
Let us assume that we have chosen a physical model of the device. Hence, we have
a collection of governing parameters of the model uniquely determining the state of the device.
Such governing parameters are for example boundary conditions, material parameters or geometrical 
features. If we assign a value to each parameter, the values of all terms from Equation \ref{eq:sysEnBalEp1}
are accessible, in principle, by means of computation. Hence the electric power, entropy production
and all energy fluxes through the boundary are determined by the governing parameters through the chosen physical
model.

Fixing the boundary energy flux value is easy when the
energy flux is considered as one of the governing parameters. Otherwise, the boundary energy flux
value depends on the governing parameters and is determined by the chosen model.
In such case fixing the flux value generally means that not all values of the governing parameters
are suitable. Respecting such constraint requires to distinguish some parameters as
dependent and adjust their value in order to satisfy the constraint. These parameters, values of which 
are being changed during the optimization, are referred to as optimization parameters while the fixed
parameters will be called governing.

Optimization means adjusting an optimization parameter in such a way that a cost or profit functional is minimized
or maximized, respectively. Let us restrict us to case of maximizing the electric power. 
The electric power is, in principle, unbounded, so that we need to assume a constraint on energy resources flowing through the boundary. Assumption of constrained power sources is in this case inevitable, which means that at least
one optimization parameter is needed.

\section{Simple solid oxide fuel cell model}
\label{sec:model}
A concrete example of fuel cell optimization is shown in this section, and
validity of EPM hypothesis is examined. 
The model is chosen and simplified so that it is analytically tractable.
The purpose of the model is not to describe a real device, but to elucidate the relations between optimization and EPM.
A variety of optimization constraints is tried out in order to expose
the limits of EPM.
\begin{figure}
\centering
\includegraphics[scale=1.0]{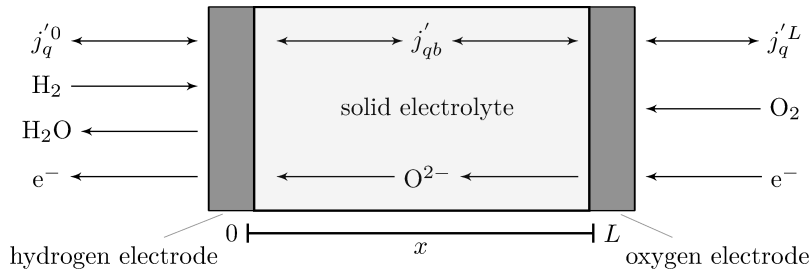}
\caption{Scheme of the one-dimensional solid oxide fuel cell model. 
The cell consists of 3 parts - the HOR surface at $x=0$, the electrolyte inbetween $(0,L)$, and 
the ORR surface at $x=L$. The surfaces are considered as points in the one-dimensional model, 
they are assumed infinitely thin. Quantities in the electrolyte are often denoted by subscript $b$ as bulk. 
Current density is constant thorough the cell and is denoted by $j$ {$(\rm A\, m^{-2})$, for definition see
footnotes \ref{note3} and \ref{note1}}.
Temperature is considered continuous in $[0,L]$.}
\label{fig:cell}
\end{figure}
\subsection{Solid oxide fuel cell}
Considered a solid oxide fuel cell composed of three parts as it is illustrated in figure 
\ref{fig:cell}. The ionic conductive solid is enclosed by two reaction surfaces, where oxygen 
reduction and hydrogen oxidation, respectively, take place. The fuel cell model works as follows.
Oxygen molecules on the right reaction surface enter the reaction while decomposing 
and accepting electrons. Then, the ions formed
on the reaction surface are transported through the solid. Finally, the ions are stripped of the electrons and form
water vapor in reaction with hydrogen on the left reaction surface. The electrochemical reactions read
\begin{eqnarray}%
\label{eq:orr}\cee{\frac{1}{2}O_2 + 2e^- \longrightarrow &O^{2-}}\quad &\mbox{at}\quad x = L,\\ %
\label{eq:hor}\cee{H_2 + O^{2-} \longrightarrow &H_2O + {2e^-}}\quad &\mbox{at}\quad  x=0,%
\end{eqnarray}%
Electrons produced in hydrogen oxidation reaction (HOR) flow through an outer circuit, where load -- for example a
resistor -- is connected, to the surface where oxygen reduction reaction takes place (ORR). Both reactions, HOR and ORR, can be added to the overall reaction
\begin{equation}
\label{eq:totalChem}\cee{H_2 + \frac{1}{2}O_2 \longrightarrow  H_2O}, %
\end{equation}
chemical (Gibbs) energy of which is being converted to electrical work.

\subsubsection*{Solid electrolyte}
This article is restricted to a steady state in one-dimensional approximation for the sake of simplicity.
The considered electrolyte consists of a segment $(0,L)$ of a conductive solid subject to boundary conditions.
Because there are no electrochemical reactions taking place in the electrolyte, the oxygen ions are neither consumed nor created therein, and flux of the ions is thus constant in space (due to the 1D approximation), 
\begin{equation}%
\dd{j}{x} = 0.%
\end{equation}%
Total electric current in the electrolyte is given by electric current of the ions, i.e. $j=i_{\rm i} = z_{\rm i} F j_{\rm i}$.
The electric potential of oxygen ions is defined analogously to electric potential of
electrons in equation \eqref{eq:electron_potential}, and entropy flux due to the ions is included into the bulk measurable 
heat flux, 
\begin{equation}\label{eq:heatFlElec}
\jqb = \jq + T\frac{j}{z_{\rm i} F}s_{\rm i},
\end{equation}
which expresses heat transport within the bulk solid electrolyte. 
The total energy density balance \eqref{eq:sysEnBal1} then becomes
\begin{equation}%
\label{eq:total_en_b}%
0 = \dd{j_{\rm en}}{x} = \dd{}{x}(j_{\rm qb}' +\phi_{\rm i} j)\ \implies -\dd{j_{\rm qb}'}{x} = j \dd{\phi_{\rm i}}{x}.\\%
\end{equation}%

In general, the stationary entropy density balance reads
\begin{equation}
\frac{\partial\js}{\partial x} = \sigmas,
\end{equation}
where $\sigmas$ and $\js$ stands for entropy production density and total entropy flux, respectively.
Inside the solid electrolyte the stationary entropy {\bf density} balance is as follows 
\begin{eqnarray}\label{eq:entropy_production_density}%
\dd{}{x}\dfrac{j_{\rm qb}'}{T} =%
{- \dfrac{j_{\rm qb}'}{T^2}\dd{T}{x} - \frac{j}{T}\dd{\phi_{\rm i}}{x}},%
\end{eqnarray}%
where the left hand side is derivative of entropy density flux and the right hand side is
the entropy production due to transport of heat and ions.
The force flux relations read
\begin{eqnarray}%
\label{eq:nablaT}\lambda\dd{T}{x}	= -j_{\rm qb}' 					+ \frac{TS^*}{F}{ j},\\%
\label{eq:nablaPhi}\dd{\phi_{\rm i}}{x}		= -\frac{S^*}{F}\dd{T}{x} 	- r{ j},%
\end{eqnarray}%
where $j_{\rm qb}'$, ${ j}$, $T$ and $\phi_{\rm i}$ stand for measurable heat flux, electric current, temperature
and electric potential of the ions, respectively, see \cite{Kjelstrup}, Eqs. 9.8. The $r$ is an electric resistivity, 
$\lambda$ is thermal conductivity at zero current, and $S^*$ stands for the transported
entropy, which is, in accordance with \cite{Kjelstrup} Eq. 9.6, defined as
\begin{eqnarray}
\pi = TS^* = \left(\frac{j_{\rm qb}'}{{ j}/F}\right)_{T = \text{const.}},
\end{eqnarray}
where $\pi$ is Peltier coefficient.
These equations describe interaction between charged species and %
temperature gradient.

Straightforward integration of equation \eqref{eq:nablaPhi} gives that 
\begin{equation}\label{eq:tok}
\Delta\phi_{\rm i} = -\frac{S^*}{F}\Delta T - rjL.
\end{equation}
Differentiating equation (\ref{eq:nablaT}) with respect to $x$, consequently 
introducing the total energy balance (\ref{eq:total_en_b}) in order to get rid of the 
$\dd{}{x}j'_{\rm qb}$ term and comparing the results with equation (\ref{eq:nablaPhi})
multiplied by the $j$ yields
\begin{equation} 
\lambda\frac{\rm d^2}{{\rm d}x^2}T = - rj^2.
\end{equation}
This last equation is a linear differential second-order one, and equipped with the boundary temperatures it gives
\begin{equation}\label{eq:temp}
T(x) = -\frac{rj^2}{2 \lambda}x^2 + \Big(\frac{\Delta T}{L} + \frac{rj^2L}{2\lambda}\Big)x + T^0,\quad x \in [0,L].
\end{equation}
Consequently, from equation (\ref{eq:nablaT}) the measurable heat flux becomes
\begin{eqnarray}\label{eq:heatfl}
j_{\rm qb}'(x) =  rj^2 x - \lambda\frac{\Delta T}{L} - \frac{rj^2L}{2} +\frac{S^*j}{F}T(x),\quad x \in [0,L].
\end{eqnarray}

Integrating the local entropy production density in the solid electrolyte,
(\ref{eq:entropy_production_density}), expressed in the terms of temperature gradient and current along
the the bulk of the electrolyte yields
\begin{equation}\label{eq:integral}
\Pi_{\rm b} = \int_0^L \sigmas = \int_0^L \frac{\lambda}{T^2}\Big(\frac{{\rm d}T}{{\rm d}x}\Big)^2 +  \frac{rj^2}{T^2}\,{\rm d}x.
\end{equation}
It is possible to evaluate this integral analytically due to the quadratic behavior of temperature. 
Thus, using explicit temperature formula (\ref{eq:temp}), total entropy production inside the electrolyte becomes 
\begin{equation}\label{eq:entropy_production}
\Pi_{\rm b} = j^2\frac{(T^L+T^0)}{2T^0T^L}rL + (\Delta T)^2\dfrac{\lambda}{T^0T^L L}.
\end{equation}
The entropy balance for the electrolyte reads
\begin{equation}\label{eq:entropy_balans}
\frac{j_{\rm{qb}}'^L}{T^L} - \frac{j_{\rm{qb}}^0}{T^0} = \Pi_{\rm b},
\end{equation}
where the boundary heat fluxes are evaluated at $0+$ and $L-$.

\subsubsection*{Surface balances}
The one-dimensional fuel cell model consists of three parts - two one-point surfaces, where reactions take place, and a bulk electrolyte. The model was solved analytically within the electrolyte in the preceding section. Let us now consider balance laws on the surfaces.
\subsubsection*{Oxygen electrode $x=L$}
Total energy balance on the surface is simply
a comparison of the energy fluxes flowing into and out of the surface. Due to the definition of measurable heat
flux on the boundary, \eqref{eq:sysEnBal2}, and measurable heat flux in the electrolyte, \eqref{eq:heatFlElec},
we observe a measurable heat flux discontinuity. This is displayed in the total energy balance of the oxygen reaction surface
as follows\footnote{\label{note3} 
From here total current density $j$ ($\rm A\,m^{-2}$) denotes both, ionic and electron, 
current densities, see footnote \ref{note1}.
Both current densities have the same value due to the one-dimensional 
assumption, nonexistent charge sources inside the fuel cell and steady-state assumption. 
Electrons are present only at the fuel cell boundary and ions are present only inside the fuel cell, therefore
there is no unambiguity.}%
\begin{equation}
\label{eq:energy_balance_L}%
j_{\rm q}'^L - j_{\rm qb}'^L- j\phi_i^L  + j\phi_e^L + j_{\rm o}^L h_{\rm o}^L=0,\quad x=L.%
\end{equation}%
Unlike the measurable heat flux, the other energy fluxes do not have their counterparts because they do not
appear on the respective sides of the surface. 

Entropy flux through the surface from the side of the electrolyte (at $L-$) is given only by the respective measurable heat flux divided by temperature at $L$, $T^L$. Entropy flux from the outer part of the surface consists of a measurable heat flux contribution and flux of entropy due to oxygen. Entropy production within the surface is given by entropy production due to the electrochemical reaction taking place therein. 
Entropy balance of the surface then reads
\begin{equation}
 \frac{j_{\rm q}'^L}{T^L} - \frac{j_{\rm qb}'^L}{T^L} + j_{\rm o}^L s_{\rm o}^L= \frac{1}{T^L}\Aft^L\,\dksi^L,%
\quad x=L,\label{eq:entropy_surf_L}
\end{equation}
where $s_{\rm o}$, $\Aft^L$
and $\dksi^L$ denote partial oxygen entropy, reaction electrochemical affinity and 
surface reaction rate, respectively. The right hand side of equation \eqref{eq:entropy_surf_L} is 
entropy production due to the surface reaction, which is the only source of entropy production on
the surface. The electrochemical affinity of oxygen reduction reaction \eqref{eq:orr} reads
\begin{equation}\label{eq:affL}
\Aft^L = 2F(\phi_{\rm i}^L-\phi_{\rm e}^L) + \dfrac{\mu_o^L}{2},\quad x=L,%
\end{equation}
in accordance with Eq. \eqref{eq:defAft}.

From the steady-state assumption and charge conservation it follows that 
\begin{equation}\label{eq:ORRrate_L}
2F\dksi^L = j,
\end{equation}
where $j$ is the electric current due to transport of the ions, also equal to $-4Fj_{\rm o}$ with $j_{\rm o}$ being molar flux of oxygen.

For simplicity we assume linear relation between electrochemical affinity $\Aft^L$ and reaction rate $\dksi^L$,
\begin{equation}\label{eq:ORRlin_L}
j = \frac{K_{\rm o}}{2RT^L}\Aft^L
\quad x=L.%
\end{equation}
The ORR current exchange density $K_{\rm o}$ is assumed to be a temperature-independent constant characterizing
kinetics of the reaction. $R$ is the universal gas constant.

Finally, combining equation \eqref{eq:ORRrate_L} with equation \eqref{eq:ORRlin_L} leads us to formula for 
the surface entropy production due to ORR, 
\begin{equation}\label{eq:entPSL}
\frac{1}{T^L}\Aft^L\,\dksi^L = \frac{R}{FK_{\rm o}}j^2.
\end{equation}

\subsubsection*{Hydrogen electrode $x=0$}
Description of situation at the HOR surface is analogous to the ORR surface. The energy
balance reads
\begin{equation}\label{eq:enBalS0}
-j_{\rm q}'^0 + j_{\rm qb}'^0 + j\phi_i^0  - j\phi_e^0 - j_{\rm h}^0h_{\rm h}^0 - j_{\rm w}^0h_{\rm w}^0=0,\quad x=0,%
\end{equation}
where we experience a similar discontinuity of the measurable heat flux as in the previous case.
The steady state condition implies $j = 2Fj_{\rm h}$ and $j = -2Fj_{\rm w}$.

The entropy balance is also analogous to the previous case,
\begin{equation}
- \frac{j_{\rm q}'^0}{T^0} + \frac{j_{\rm qb}'^0}{T^0}%
 - j_{\rm h}^0 s_{\rm h}^0 - j_{\rm w}^0 s_{\rm w}^0= \frac{1}{T^0}\Aft^0\,\dksi^0,%
\quad x=0,\label{eq:entBalS0}
\end{equation}

The electrochemical affinity of the HOR reads
\begin{equation}\label{eq:aff0}
\Aft^0 = 2F(\phii^0 - \phe^0) + \mu^0_{\rm h} - \mu^0_{\rm w},
\end{equation}
and charge conservation implies
\begin{equation}
2F\dksi^0 = j.
\end{equation}
As in the case of the ORR surface we assume 
a linear dependence of reaction rate on the electrochemical affinity,
\begin{equation}\label{eq:ORRlin_0}
j = \frac{K_{\rm h}}{2RT^0}\Aft^0,
\quad x=0.%
\end{equation}
where ${K_{\rm h}}$ is current exchange density characterizing the HOR kinetics.

Finally, the entropy production due the HOR at the surface is
\begin{equation}\label{eq:entPS0}
\frac{1}{T^0}\Aft^0\,\dksi^0 = \frac{R}{FK_{\rm h}}j^2.
\end{equation}
\subsection{Total entropy production}
To obtain the total entropy production of whole fuel cell model we simply
add the production in solid electrolyte, \eqref{eq:entropy_production} and productions due to
the electrochemical reactions, \eqref{eq:entPSL} and \eqref{eq:entPS0}. We obtain
\begin{multline}\label{eq:modelEntProd}
\Pitot = \Pi_{\rm b} + \frac{1}{T^0}\Aft^0\,\dksi^0 + \frac{1}{T^L}\Aft^L\,\dksi^L 
=j^2\Big(\frac{(T^L+T^0)}{2T^0T^L}rL + \frac{R(\Ko + \Kh)}{F\Ko\Kh}\Big)+ (\Delta T)^2\dfrac{\lambda}{T^0 T^L L}.
\end{multline}
\subsection{Current and voltage}
Observing affinity of the total fuel cell reaction, \eqref{eq:totalChem}, 
\begin{equation}\label{eq:Aftot}
\Aftot = \mu_{\rm h}^0 + \frac{1}{2}\mu_{\rm o}^L - \mu_{\rm w}^0,
\end{equation}
we see that it can be expressed in terms of ORR and HOR affinities from equations, \eqref{eq:affL} and \eqref{eq:aff0}, as
\begin{equation}
2Rj\big(\frac{T^L}{\Ko} + \frac{T^0}{\Kh} \big)=\Aft^L +\Aft^0 =2F\Delta(\phii - \phe) + \Aftot.
\end{equation}
Introducing the equation \eqref{eq:tok} instead of $\Delta\phii$, expressing $\Delta\phe$ yields
\begin{equation}\label{eq:IV}
\Delta\phe = \frac{\Aftot}{2F} - j\Big(rL + \frac{R}{F}\big(\frac{T^L}{\Ko} + \frac{T^0}{\Kh} \big)\Big)%
- \frac{S^*}{F}\Delta T.
\end{equation}

\section{Fuel cell model optimization}

In the preceding section we have outlined a simplified one-dimensional steady-state model of a solid oxide fuel cell.
The model was simple enough to admit analytical solution, which will be advantageous in the present section.

Before proceeding with optimization of the solid oxide fuel cell model from section \ref{sec:model}, we need to 
specify governing and optimized parameters, constraints and an objective, as we have discussed
in general in the section \ref{subsec:OptGen}. Those specifications have to respect
the physical nature of problem as well as they must neither over- nor underdetermine the model
equations.

Let us choose the electrical power $\dWel$ to be the optimization objective, which we want to maximize 
with respect to optimized parameter. 

\subsection{Optimization without a priori constraints}\label{sec:noconst}
Optimization can proceed so that all but one necessary boundary conditions are fixed and the remaining one is varied in order to attain maximum power.
\subsubsection*{Optimization of thickness L}
Let us choose material parameters $\lambda, r, S^*, \Ko, \Kh$, boundary conditions $T^L, T^0, p^L, p^0, \Delta\phe$
to be some given parameters (governing parameters) while thickness $L$ will be the optimized parameter within 
$0 < a \leq L \leq b < \infty$. In other words, we seek the thickness $L$ for which the electrical power is maximal.

Such choice of governing parameters reveals that the IV-formula \eqref{eq:IV} determines current $j$
as a decreasing function of thickness. Therefore, by definition of electric power \eqref{eq:sysEnBal2}, it follows 
that the power is maximal for the smallest possible $L$, therefore 
\begin{equation}
a = \argmax_L\dWel(L) = \argmax_L j(L),
\end{equation}
and that $\dWel$ is monotone with respect to $L$, which also means that
\begin{equation}
\max_L \dWel \leq \dWel(0).
\end{equation}
The power is thus bounded with respect to $L$, and it decreases as $L$ increases.

Let us now inspect the entropy production dependence on $L$. The values of 
entropy production \eqref{eq:modelEntProd}  tend to infinity for $L\rightarrow 0$.
Moreover, entropy production is a smooth non-negative function, i.e.
\begin{equation}
0 < \argmin_L \Pitot(L).
\end{equation}
Therefore, for $a$ sufficiently small we have
\begin{equation}
a < \argmin_L \Pitot(L)\quad\implies\quad \argmax_L \dWel(L) < \argmin_L \Pitot(L),
\end{equation}
which means that maximum power is not attained for the same $L$ as minimum entropy production.

In summary, when thickness $L$ is the parameter that is varied in order to find maximum power, maximum power is attained for smallest possible $L$. On the other hand, entropy production density tends to infinity as $L\rightarrow 0$, which means that EPM is not a valid optimization strategy in this case.

\subsubsection*{Optimization of heat conductivity $\lambda$}
In this case we assume that $L$ is a governing parameter, but thermal conductivity $\lambda$ is
the optimization parameter. The current \eqref{eq:IV} is constant with respect to $\lambda$ and so is
the electric power, see Eq. \eqref{eq:sysEnBal2}. On the contrary, entropy production \eqref{eq:modelEntProd} is increasing
with $\lambda$ increasing. Therefore, EPM does not coincide with electrical power maximization in this case. 

\subsection{Optimization with constrained resources}
It was demonstrated in the a priori unconstrained optimization examples in the preceding section that EPM often does not correspond with maximization of useful power. The examples, however, are somewhat ill-posed because
we maximize the power without paying attention to the amount of resources used. In practice, the energy resources are 
limited, therefore, we introduce constraints on "source" energy fluxes in this section.

Expressing the general formula for power \eqref{eq:sysEnBalEp1} in the particular situation
of the solid oxide fuel cell model yields
\begin{equation}\label{eq:power2}
{\dWel} =%
            \underbrace{-\jqp^L\big(1 - \frac{T^0}{T^L} \big)%
			+ \frac{j}{2F}\Big(\Aftot%
            + {\Delta T s_{\rm o}^L/2}\Big)}_\text{no a priori sign}%
            -\underbrace{\vphantom{\dfrac{T}{F^L}}T^0\Pitot}_{\geq 0},
\end{equation}
which can be rewritten as
\begin{equation}\label{eq:power3}
{\dWel} =%
            C   -T^0\Pitot\text{\quad with respect to optimization parameters.}
\end{equation}
The terms with no a priori sign are the exergy flux into the fuel cell, and if they are kept constant (denoted by $C$), 
maximization of useful power corresponds to minimization of entropy production. EPM is then a valid optimization
method in that case. 

What if we do not wish to fix all the terms with no a priori sign in equation \eqref{eq:power2}?
Does then EPM still lead to the maximum power?
The answer is negative in general, as we have shown in Sec. \ref{sec:noconst}. 
Moreover, we show in the following that even if constraints on energy influx are chosen,
EPM does not often lead to the maximum power anyway.

It is convenient to assume that the fuel cell model is connected to an external load with resistance\footnote{With a little abuse of notation, we can set $Z$ negative, which corresponds to an external voltage source.}. Using the Ohm's law,
~$\Delta\phe~=~Zj$,~for the external load, Eq. \eqref{eq:IV} yields an useful formula electric current 
\begin{equation}\label{eq:current1}
j = \dfrac{\frac{\Aftot}{2F} - \frac{S^*}{F}\Delta T}{rL + Z + \frac{R}{F}\big(\frac{T_L}{\Ko} + \frac{T_0}{\Ko} \big)}.
\end{equation}
The electric power of the fuel cell model can be also expressed as
\begin{equation}\label{eq:power1}
\dWel = Zj^2
\end{equation}
alternatively.

For all further optimization examples we assume that the set of governing parameters (that remain constant) consists
of $T^L$, $T^0$, $p_{\rm w}$, $p_{\rm o}$, $p_{\rm h}$ , $\lambda$, $r$, $\Ko$, and $\Kh$ while $Z$ and $L$ are the 
optimization parameters. It is necessary from the mathematical point of view to have two optimization parameters instead
of one because otherwise we could not enforce any constraint.

\subsubsection*{Gibbs energy flux constraint}\label{sec:GibbsConst}
Optimization with Gibbs energy flux into the fuel cell as the constraint is a natural choice in fuel cells because efficiency is often expressed as 
\begin{equation}
 \eta = \frac{\dWel}{\Aftot\frac{j}{2F}},
\end{equation}
where the denominator is just the Gibbs energy flux. Hence, we consider 
\begin{equation}\label{eq:gibbsConstr}
\frac{j}{2F}\Aftot(Z,L) = C \text{,\quad $C$ is constant.}
\end{equation}

Total affinity, given by formula \eqref{eq:Aftot}, is independent of $Z$ and $L$, and constraint \eqref{eq:gibbsConstr} in fact 
fixes the current $j$. Observing the formula for $j$, Eq. \eqref{eq:current1}, we see that
constraint \eqref{eq:gibbsConstr} implies 
\begin{equation}\label{eq:ZCrL}
Z = C_\text{ref} - rL\text{,\quad where $C_\text{ref}$ is a positive constant.}
\end{equation}
Both optimization parameters, $Z$ and $L$, have to be positive. Assuming that $L$ has to be
greater than some smallest possible positive thickness $a$, we see that maximum value of Z is
\begin{equation}\label{eq:GibbsWMax}
Z_\text{max} = C_\text{ref} - ra.
\end{equation}
Because $j$ is constant, electric power is linear in $Z$, and maximum power is achieved
when $Z$ is maximal possible, hence,
\begin{equation}
\argmax_Z \dWel = C_\text{ref} - ra\text.
\end{equation}
This last equation identifies the value of $Z$ for which the power is maximal.

Let us now search for minimum of entropy production. Introducing relations \eqref{eq:ZCrL} into entropy production \eqref{eq:modelEntProd} gives 
\begin{equation}\label{eq:PitotGibbsConstr}
\Pitot  = j^2 \left(\frac{(T^L+T^0)}{2T^0T^L}(C_\text{ref}-Z) + \frac{R(\Ko + \Kh)}{F\Ko\Kh}\right)+ \dfrac{(\Delta T)^2\lambda r}{T^0T^L (C_\text{ref}-Z)},
\end{equation}
minimum of which is (given by solving a quadratic equation)
\begin{equation}\label{eq:GibbsPiMin}
\argmin_Z \Pitot = C_\text{ref} - \left|\frac{\Delta T}{j}\right|\left(\frac{2\lambda r}{T^L + T^0} \right)^\frac{1}{2}.
\end{equation}
In general we have
\begin{equation}\label{eq:GibbsConst.result}
\left[ ra \neq \left|\frac{\Delta T}{j}\right|\left(\frac{2\lambda r}{T^L + T^0} \right)^\frac{1}{2}\right]%
\implies%
\left[\argmin_Z \Pitot \neq\argmax_Z \dWel\right]
\end{equation}
Electric power thus gains maximum value for different $Z$ than at which entropy production reaches minimum.
Optimization with constrained Gibbs energy flux is demonstrated in figure \ref{fig:plotG}.

It was shown in \cite{GenExAl} that maximum of electric power coincides with minimum of entropy production if $\Delta T =0$, which can be seen also from Eq. \eqref{eq:power2} easily. How is this result reflected in Eq. \eqref{eq:GibbsConst.result}?
Entropy production \eqref{eq:modelEntProd} becomes linear in $L$ in the isothermal case and is thus minimal when $Z$ is maximal, 
i.e. where $\dWel$ is maximal. In formula \eqref{eq:GibbsConst.result} the left hand side becomes zero as well as the right hand side. This also agrees with Fig. \ref{fig:plotG}, where extremal values $\Pitot$ and $\dWel$ tend to each other as $\Delta T\rightarrow 0$. 

In summary, when flux of Gibbs energy is kept constant during the optimization, useful power and entropy production do not reach extrema (maximum and minimum, respectively) simultaneously if the fuel cell is not isothermal. If the fuel cell is isothermal, the extrema coincide.
\subsubsection*{Heat and Gibbs energy flux constraint}
\begin{figure}%
\centering
	\begin{subfigure}{\textwidth}
	\centering
	\includegraphics[scale=1.0]{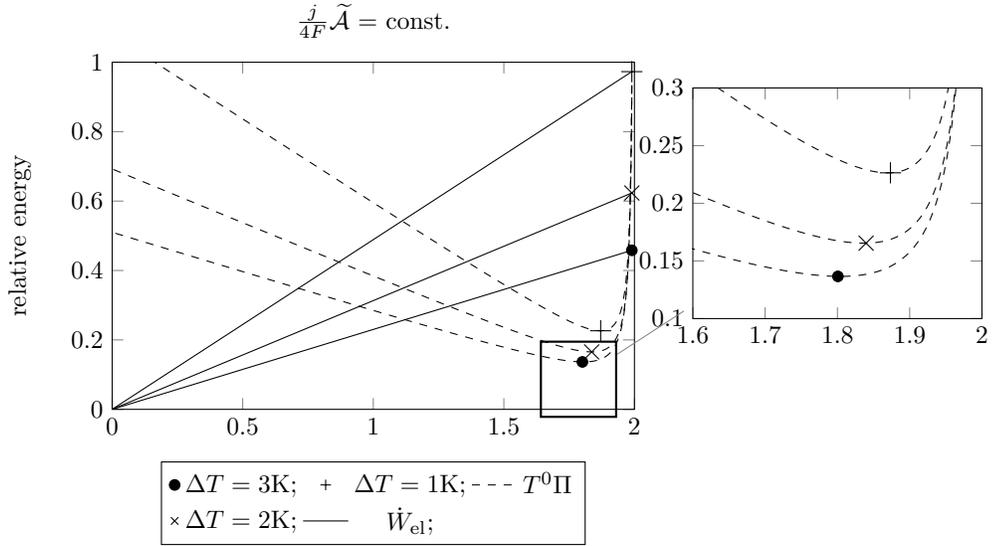}
		\caption{Gibbs energy flux is kept constant. The maxima of power are attained on boundary for
		every temperature difference. The entropy minima tend to power maxima with decreasing temperature
		difference.}
		\label{fig:plotG}
	\end{subfigure}
	\begin{subfigure}{\textwidth}
	\centering
	\includegraphics[scale=1.0]{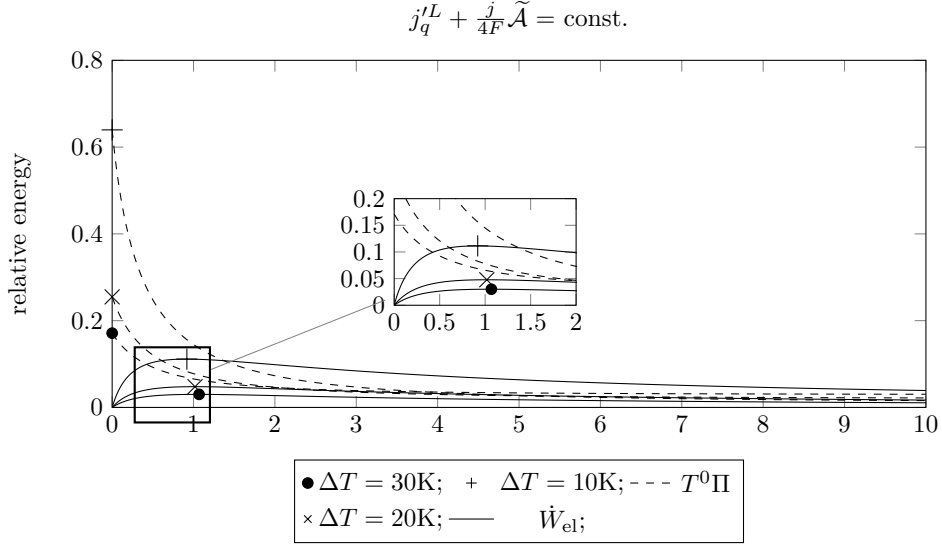}
		\caption{Sum of Gibbs energy flux and heat from hot reservoir is constrained. 
		Entropy production minima lie at boundary $Z=10$. Work maxima tend move to $Z=0$, with decreasing
		temperature difference.}
		\label{fig:plotGQ}
	\end{subfigure}
\caption{Marks denote extremes of power and entropy production, respectively. Both, entropy production
	and power values are relative to their value at $[Z,L] = [1,1]$. Boundary temperature $T^L$ is in
	every case set constant to $1073K$, $T^0$ is varied.}
\end{figure}

Another example of constraint is to fix both Gibbs energy flux and heat flux through the hot reservoir, assuming $T^L\ge T^0$,
\begin{equation}\label{eq:ghConstr}
-\jqp^L(Z,L) + \frac{j(Z,L)}{2F}\Aftot = C \text{,\quad $C$ is constant.}
\end{equation}
The implicit relation which binds $Z$ and $L$ is no longer as simple as in the case of Gibbs energy flux constraint, Sec. \ref{sec:GibbsConst}.
Unlike as in the previous case, we cannot write an explicit formula relating $Z$ and $L$. Nevertheless, plugging constraint
\eqref{eq:ghConstr} into equation \eqref{eq:power2}, we obtain
\begin{equation}\label{eq:power4}
{\dWel} =%
            \underbrace{-\jqp^L + \frac{j}{2F}\Aftot}_\text{constant} %
			+\frac{T^0}{T^L}\jqp^L
			+ \frac{j}{2F}\left(\Aftot + {\Delta T s_{\rm o}^L/2}\right) 
            -\underbrace{\vphantom{\dfrac{T}{F^L}}T^0\Pitot}_{\geq 0}.
\end{equation}
The non-constant terms in front of the entropy production in this formula make EPM invalid also in this case. 
The corresponding electric power and entropy production are plotted in figure \ref{fig:plotGQ} for different boundary temperatures.

\subsection{Exergy flux as constraint}
Finally, one can consider 
\begin{equation}\label{eq:TghsConst}
-\jqp^L\left(1 - \frac{T^0}{T^L} \right)%
+ \frac{j}{2F}\left(\Aftot + {\Delta T s_{\rm o}^L/2}\right) 
 = C \text{,\quad $C$ is constant.}
\end{equation}
Such constraint collapses equation \eqref{eq:power2} into the form of equation \eqref{eq:power3}, and 
minimum of entropy production thus implies maximum of power for any couple of optimization parameters in this case. Constraint \eqref{eq:TghsConst} however coincides with exergy flux into the fuel cell.

Instead of fixing the whole flux of exergy into the fuel cell, we can fix both its components separately, i.e.
\begin{equation}\label{eq:TiGSConst}
\jqp^L (Z, L, \lambda)= C_1,\quad \mbox{and } \frac{j(Z,L)}{2F}\Aftot = C_2 \text{,\quad $C_1$ 
and $C_2$ are constant.}
\end{equation}
Having two constraints, we have to work with three optimization parameters, for example $Z, L, \lambda$. 

Note that by Eq. \eqref{eq:current1} current $j$ does not depend on $\lambda$. Therefore, the second constraint
in \eqref{eq:TiGSConst} implies that the Gibbs energy flux is also fixed.
Treating $L$ as a function of $Z$, the first constraint in \eqref{eq:TiGSConst}
yields the following dependence of $\lambda$ on $Z$:
\begin{equation}\label{eq:jqConstr}
\lambda = \dfrac{C_{\rm ref}-Z}{r\Delta T}\left[-C_1 + j\frac{T^L}{F}\left(S^* + \frac{s_{\rm o}}{4} \right) 
+ j^2\left(\frac{C_{\rm ref}-Z}{2}+  \frac{RT^L}{F\Ko}\right)\right].
\end{equation}
Finally, we plug \eqref{eq:jqConstr} into entropy production \eqref{eq:modelEntProd}, and get
\begin{equation}
T^0\Pi(Z) = j^2\left(C_{\rm ref}-Z + \frac{R}{F}\left(\frac{T^L}{\Ko} +\frac{T^0}{\Kh}\right)\right)
+ j\frac{\Delta T}{F}\left(S^* + \frac{s_{\rm o}}{4} \right) - \left(1-\frac{T^0}{T^L}\right) C_1.
\end{equation}
Bearing in mind that $j$ is constant, we can see that the minimum of entropy production coincides with the highest possible value of
$Z$, where $Z$ is limited by the minimal thickness $a$ as in equation \eqref{eq:GibbsWMax}. Maximum of electric power
is also reached at the maximal possible value of $Z$, reasoning of which is the same as in section \ref{sec:GibbsConst}. 
This is, however, not surprising, since fixing constraints \eqref{eq:TiGSConst} inevitably leads to fixed flux of exergy into the fuel cell.
\clearpage

\section{Conclusion}
Efficiency of industrial devices producing electricity is often examined by means of exergy analysis, that means by evaluating exergy destruction within the devices. Since exergy destruction is proportional to entropy production, reducing exergy destruction in fact means reducing entropy production. But the final goal of such optimization is to raise the useful (electric) power delivered by the device. Is reduction of entropy production always accompanied by growth of the useful power? Not in general, as is demonstrated in this paper on several examples.

Before saying whether entropy production minimization (EPM) leads to useful power maximization in a particular case, it is necessary to state what are the constraints of the optimization, i.e. which quantities are kept fixed. For example when exergy flux into the device is fixed (either as a whole or each part of it), EPM is equivalent to maximization of useful power. Similarly when the maximum work is sought that a device can deliver when relaxing to equilibrium (being shut down), the maximum is obtained when no entropy is produced.

Consider now a fuel cell, which can be seen as a prototype of device converting chemical energy into electricity. Therefore, it is reasonable to keep only flux of Gibbs energy into the fuel cell constant during the maximization. Indeed, Gibbs energy expresses the useful energy of the fuel while flux of exergy also contains heat fluxes from all but one temperature reservoirs. If flux of Gibbs energy into the fuel cell is fixed and boundary of the fuel cell is isothermal, then EPM again leads to useful power maximization. On the other hand, if the boundary is not isothermal, EPM fails to provide maximum useful power, see Fig. \ref{fig:plotG}. The situation is similar when flux of Gibbs energy and the heat flux from the hotter temperature reservoir are kept fixed as useful power and entropy production attain their respective extrema at different conditions, see Fig. \ref{fig:plotGQ}. Finally, not fixing any energy flux through the boundary makes EPM also inadequate for useful work maximization.

In summary, before assessing efficiency of a device by means of exergy analysis, one should first define the optimization procedure, which includes defining constraints fixed during the optimization, and then one should verify that entropy production minimization is equivalent to useful power maximization in the particular case given by the device and the optimization procedure. Skipping any of these steps, one may end up in a pitfall hidden behind the widely used theory of exergy analysis.
%
%
\section*{Acknowledgement}
We are grateful to Miroslav Grmela, who supported this research. We are also grateful to Václav Klika for encouraging us and discussing the results.
%
The work was partially developed within the POLYMEM project, under registration number~CZ.1.07/2.3.00/20.0107, that is co-funded from the European Social Fund (ESF) in the Czech Republic: "Education for Competitiveness Operational Programme", from the CENTEM project, reg. no. CZ.1.05/2.1.00/03.0088, cofunded by the ERDF as part of the Ministry of Education, Youth and Sports OP RDI programme and, in the follow-up sustainability stage, supported through CENTEM PLUS (LO1402) by financial means from the Ministry of Education, Youth and Sports under the ”National Sustainability Programme I.``.

%
The work was supported by Czech Science Foundation (project no. 14-18938S).

The study was supported by the Charles University in Prague, project GA UK No 70515.
%

%
\end{document}